\documentstyle[epsf,epsfig]{aipproc}
\begin{document}
%%%%%%%%%%%%%%%%%%%%%%%%%%%%%%%%%%%%%%%%%%%%%%%%%%%%%%%%%%%%%%%%%%%%%%%%%%%%%%%%
%%%%%%%%%%%%%%%%%%%%%%%%%%%%%%%%%%%%%%%%%%%%%%%%%%%%%%%%%%%%%%%%%%%%%%%%%%%%%%%%
\newcommand {\bay } {\begin{array}   } 
\newcommand {\eay } {\end{array}   }
\newcommand {\ol  } {\overline       } 
\newcommand {\lb }  {\label }
\newcommand {\slL}  {\sum_{\ell=0}^L                    }
\newcommand {\nl }  { \newline                           }
\newcommand {\cF  } {{\cal   F }}       
\newcommand {\cH  } {{\cal   H }}       
\newcommand {\cK  } {{\cal   K }}       
\newcommand {\cN  } {{\cal   N }}      
\newcommand {\cZ  } {{\cal   Z }}      
\newcommand {\hsc } {\hspace*{1cm}   }
\newcommand {\fa  } {\forall                         }
\newcommand {\lh  } {\left(                          }
\newcommand {\rh  } {\right)                         }
\newcommand {\lv  } {\left[                          }
\newcommand {\rv  } {\right]                         }
\newcommand {\lgl } {\langle                    }
\newcommand {\rgl } {\rangle                   }
\newcommand {\lc  } {\left\{                         }
\newcommand {\rp  } {\right .                        }
\newcommand {\si  } {\sigma     }
\newcommand {\be  } {\beta      }
\newcommand {\de  } {\delta     } 
\newcommand {\ep  } {\epsilon   } 
\newcommand {\ta  } {\tau       }
\newcommand {\vsi } {{\vec{\sigma  }}}
\newcommand {\vx  } {{\vec{  x }}}
\newcommand {\Tr  } {\mathop{\mbox{\rm Tr}}   }
\newcommand {\tim } {{\tilde{ m }}}
\newcommand {\tin } {{\tilde{ n }}}
\newcommand {\ev  } {\equiv}
\newcommand {\e   } {\!+\!                  }
\newcommand {\m   } {\!-\!                  }
\newcommand {\noi } {\noindent       }
\newcommand {\ov  } {\over }
%%%%%%%%%%%%%%%%%%%%%%%%%%%%%%%%%%%%%%%%%%%%%%%%%%%%%%%%%%%%%%%%%%%%%%%%%%%%%%%%

\title{\bf Cluster Derivation of the Parisi Scheme for 
Disordered Systems}

\author{A.C.C.~Coolen and J.~van~Mourik}
\address{  Department of Mathematics, King's College,
           University of London\\
           The Strand, London WC2R 2LS, U.K.
        }
\maketitle

%%%%%%%%%%%%%%%%%%%%%%%%%%%%%%%%%%%%%
\begin{abstract}
%%%%%%%%%%%%%%%%%%%%%%%%%%%%%%%%%%%%%
We propose a general quantitative scheme in which systems are given the freedom
to sacrifice energy equi-partitioning on the relevant time-scales of 
observation, and have phase transitions by separating autonomously into ergodic 
sub-systems (clusters) with different characteristic time-scales and 
temperatures. The details of the break-up follow uniquely from the requirement 
of zero entropy for the slower cluster. Complex systems, such as the 
Sherrington-Kirkpatrick model, are found to minimise their free energy by 
spontaneously decomposing into a hierarchy of ergodically equilibrating degrees
of freedom at different (effective) temperatures. This leads exactly and 
uniquely to Parisi's replica symmetry breaking scheme.
Our approach, which is somewhat akin to an earlier one by Sompolinsky, gives new
insight into the physical interpretation of the Parisi scheme and its relations
with other approaches, numerical experiments, and short range models. 
Furthermore, our approach shows that the Parisi scheme can  be {\it derived}
quantitatively and uniquely from plausible physical principles.
\end{abstract}

%%%%%%%%%%%%%%%%%%%%%%%%%%%%%%%%%%%%%%%%%%%%%%%%%%%%%%%%%%%%%%%%%%%%%%%%%%%%%%%%
\section{Introduction \lb {IN}}
%%%%%%%%%%%%%%%%%%%%%%%%%%%%%%%%%%%%%%%%%%%%%%%%%%%%%%%%%%%%%%%%%%%%%%%%%%%%%%%%

The so-called Parisi scheme \cite{P} for replica symmetry
breaking (RSB) has been one of the most succesful tools in the description of
(the statics of) disordered systems in the non-ergodic or glassy phase. 
Originally proposed as the solution for the Sherrington-Kirkpatrick (SK)-model
\cite{SK} for mean field spin glasses, it has been succesfully applied to a
wide range of disordered systems. The physical interpretation  of the Parisi 
scheme has been a subject of many discussions, and has led to the introduction
of concepts such as {\it disparate time-scales} \cite{S}, {\it effective 
temperatures} \cite{BP}, {\it low entropy production} \cite{CK} and 
non-equilibrium {\it thermodynamics} \cite{N}.

In this paper, we will show how a general scheme in which systems are given 
the freedom to sacrifice energy equi-partitioning by separating autonomously 
into sub-systems with different characteristic time-scales and temperatures, 
not only yields the Parisi scheme, but also introduces all the above-mentioned 
concepts in a very natural way. Our assumptions are simple, and all the 
quantities that appear in the theory, have a clear physical meaning.
In section \ref{DT} we briefly discuss systems with disparate time-scales, 
and in section \ref{SK} we apply this to the benchmark problem of mean field 
disordered systems, viz. the SK-model \cite{SK}.  In section \ref{NU} we 
present numerical evidence for the existence of multiple disparate time-scales.
Finally, in the discussion \ref{DI} we summarize the simple physical picture 
that naturally emerges from our scheme, and discuss the points that still need 
further investigation.

%%%%%%%%%%%%%%%%%%%%%%%%%%%%%%%%%%%%%%%%%%%%%%%%%%%%%%%%%%%%%%%%%%%%%%%%%%%%%%%%
\section{Systems with Disparate Time Scales \lb {DT}}
%%%%%%%%%%%%%%%%%%%%%%%%%%%%%%%%%%%%%%%%%%%%%%%%%%%%%%%%%%%%%%%%%%%%%%%%%%%%%%%%

In this section we briefly describe how the formalism, as developed and applied
for Ising spin systems with slowly evolving bonds \cite{PC}, can be generalised
to arbitrary stochastic systems with two or more disparate time-scales. In the 
case of systems with two infinitely disparate time-scales, the faster variables
$\vx _f$ equilibrate before the slower ones $\vx _s$ can effectively change.
Therefore, the $\vx _s$ evolve to an a Boltzmann-type equilibrium distribution
with an effective energy which is the free energy of the $\vx _f$ given the
$\vx _s$, at an effective inverse temperature $\be _s$, while the $\vx _f$
evolve to the normal Boltzmann equilibrium distribution with $\be _f=\be$:

\begin{equation}
\bay {llll}
&\cZ &=&\Tr _{\vx _s}~\lh \cZ _f(\vx _s)\rh ^\tim ~,~~\\[1mm]
&\cZ _f(\vx _s)&\ev &\Tr _{\vx _f}~\exp\lh -\be \cH (\vx )\rh ~.~~
\lb {dt:Zf}
\eay 
\end{equation}

\noi One thus obtains a theory with a non-negative replica dimension $\tim \ev
{\be _s/\be }$. This procedure can be generalised to a system with $L\e 1$ 
different levels of stochastic variables, time-scales and temperatures $\{(\vx
_\ell,~\ta _\ell,~\be _\ell):~\ell\in\{0,L\}~\}$. Assuming that each level is 
adiabatically slower than the next level (${\ta _\ell/\ta _{\ell-1}}=0$), we 
obtain the following recursion relations

\begin{equation}
\bay {llll}
&\cZ _\ell&\ev &\Tr _{\vx _\ell}~\lh \cZ _{\ell+1}\rh
^{\tim_{\ell+1}}~~~(\ell<L), \\[1mm]
&\cZ _L   &\ev &\Tr _{\vx _L}~\exp\lh -\be _L~\cH (\{\vx \})\rh ~,~~
\eay 
\lb {dt:Zl}
\end{equation}

\noi where $\tim_ \ell\ev \be _{\ell-1}/\be _\ell,~\be _L=\be $, and the total 
free energy of the system, defined on the longest time-scale, is given by

\begin{equation}
\cF =-{1\ov \be _0}~{\log(\cZ _0)}~.~~
\lb {dt:F}
\end{equation}

%%%%%%%%%%%%%%%%%%%%%%%%%%%%%%%%%%%%%%%%%%%%%%%%%%%%%%%%%%%%%%%%%%%%%%%%%%%%%%%%
\section{The SK-model \lb {SK}}
%%%%%%%%%%%%%%%%%%%%%%%%%%%%%%%%%%%%%%%%%%%%%%%%%%%%%%%%%%%%%%%%%%%%%%%%%%%%%%%%

We will now apply this scheme to the SK-model \cite{SK}, for which the Parisi 
scheme was originally developed. Therefore, we briefly recall the definitions:

\begin{equation}
\cH (\vsi )=-\sum_{i<j}^N J_{ij}\si _i\si _j~,
\lb {sk:H}
\end{equation}

\noi with Gaussian couplings $J_{ij}$, $P(J_{ij})=\cN(J_0/N,J/\sqrt{N})$, and
Ising spins $\si _j$.

We assume that there are $L\e 1$ levels of spins with corresponding disparate
time-scales and temperatures $\{(\vsi _\ell=\{\si _j\in I_\ell\},|I_\ell|\ev 
\ep _\ell,\ta _\ell,T_\ell),\ell=0,..,L\}$ in the system, where ${\ta _\ell/\ta
_{\ell\m 1}}=0$, such that larger $\ell$ correspond to faster spins. Although we
expect the selection of time-scales for the spins to depend on the specific
realisation of the couplings, at present we will make the simplest 
approximation: the system can only choose the relative sizes $\ep _\ell$ of the
levels. A more detailed study of the (annealed) selection of levels will be
presented elsewere. To deal with the quenched disorder average over the $J_{ij}$
we use the replica trick 

\begin{equation}
\ol {\cF }=-{1\ov \be _0}\ol {\log\cZ _0}=-\lim_{\tin \to 0}{1\ov \tin \be _0}
\log\ol {\cZ _0^\tin }~.~~
\lb {sk:F}
\end{equation}

\noi Together with the recursion relations (\ref{dt:Zl}) this leads to a nested
set of $\tin \prod_{\ell=1}^L\tim _\ell$ replicas, in which spins at level 
$\ell$ carry a set of replica indices $\{a\}_\ell\ev \{a_0,..,a_\ell\}$. The 
index $a_0=1,..,\tin $ comes from the disorder average, whereas 
$a_\ell=1,..,\tim _\ell=\be _{\ell-1}/\be _\ell$. The asymptotic free energy 
per spin is then given by

\begin{eqnarray}
f&=&\lim_{\tin \to 0}{-1\ov \tin \be _0}~\lv {-J^2\be ^2\ov 4}\!\!\sum_{\{a\}_L,
\{b\}_L} q^{\{a\}_L~2}_{\{b\}_L}+\slL ~\ep _\ell~\log(\cK _\ell)\rv ~,~~
\lb {sk:fab}\\
\cK _\ell&\ev &\Tr_{\si ^{\{c\}_\ell }}\exp\lv {J^2\be ^2\ov
2}\!\!\!\sum_{\{a\}_L, \{b\}_L}q^{\{a\}_L}_{\{b\}_L}~\si ^{\{a\}_\ell}\si ^{\{b\}_\ell}\rv ,~~~~
q^{\{a\}_L}_{\{b\}_L}\ev {1\ov N}\sum_{\ell=0}^L\sum_{j\in I_\ell}
\si _j^{\{a\}_\ell}\si _j^{\{b\}_\ell} .
\lb {sk:qab}
\end{eqnarray}

\noi With the definitions $m_\ell\ev \prod_{f=\ell}^L~\tim_ f=\be _{\ell-1}/\be
$ we have $\be _0\tin =\be n$, and the connection with the original Parisi
scheme becomes clear. Note that $0\leq\tim _\ell
\leq 1$, as slower clusters cannot have a lower temperature than faster ones,
because otherwise the latter would act as a heat bath. We now assume full
ergodicity at each level in the hierarchy of time-scales: 

\begin{equation}
q^{\{a\}_L}_{\{b\}_L}=q_\ell~,\hsc \ell\ev \sum^L_{r=0}~r~
\de _{\{a\}_{r-1},\{b\}_{r-1}}~(1-\de _{a_r,b_r})~,~~
\lb {sk:qrs} 
\end{equation}

\noi (i.e. $\ell$ is the slowest level at which $\{a\}_L$ and $\{b\}_L$ differ),
to obtain

\begin{equation}
f={J^2\be \ov 2}\slL \lv {m_{\ell\e 1}\ov 2}~(q^2_{\ell\e 1}-
q_\ell ^2)-\ep _\ell ~[q]_\ell \rv -{1\ov m_1\be }\slL ~\ep _\ell\int 
Dz_0\log(\cN ^1_\ell )~,~~~
\lb {sk:frs}
\end{equation}

\noi where

\begin{eqnarray}
   \cN ^r_\ell    &\ev& \lc \bay {lll}
\int Dz_r \lh \cN ^{r+1}_\ell\rh ^{m_r\ov m_{r\e 1}} 
&,&\hsc r\leq \ell \\
        2\cosh(J\be m_{\ell\e 1}\sum_{s=0}^\ell z_s\sqrt{q_s\m q_{s\m 1}}) 
&,&\hsc r=\ell+1         \eay \rp 
\lb {sk:N}\\
\lv q\rv _\ell&\ev &\sum_{r=\ell}^{L}m_{r\e 1}~(q_{r\e 1}-q_r)~,~~
\lb {sk:[q]}
\end{eqnarray}

\noi The physical meaning of the $q_\ell$ is given by

\begin{equation}
q_\ell=\lim_{N\to\infty}\sum_j\ol {\lgl ~..~\lgl ~\lgl ~\lgl ~..~\lgl \si _j 
\rgl _L~..~\rgl _{\ell\e1} ~\rgl _\ell^2 ~\rgl _{\ell\m 1}~..~\rgl _0}~,\hsc
\lb {ski:qm}
\end{equation}

\noi where $\lgl \cdot\rgl _r$ denotes the average over the level $r$ process, 
and $\ol {~\cdot~}$ denotes the disorder average. The minimum of the free 
energy with respect to the $\ep _\ell$ (with $\sum_{\ell=0}^L~\ep _\ell =1$, 
for $\tin , \tim _\ell$  positive integers), is at 
$\{\ep ^*_L=1,~\ep ^*_\ell=0~~\fa \ell<L\}$, and we exactly recover the $L$-th
order Parisi solution. Furthermore, the extremization with respect to the 
$\tim _\ell$ is now recognized to express the fact that, for self-consistency,
the entropy of the spins slower than level $\ell$ is zero (they are effectively
fixed on the time-scale $\ta _\ell$).

%%%%%%%%%%%%%%%%%%%%%%%%%%%%%%%%%%%%%%%%%%%%%%%%%%%%%%%%%%%%%%%%%%%%%%%%%%%%%%%%
\section{Numerical Evidence \lb {NU}}
%%%%%%%%%%%%%%%%%%%%%%%%%%%%%%%%%%%%%%%%%%%%%%%%%%%%%%%%%%%%%%%%%%%%%%%%%%%%%%%%

In numerical simulations of the SK-model we have measured the distribution of 
the number of flips $f$ at time $t$ per spin: $\rho_{\rm sim}(f,t)$. Assuming a 
characteristic time-scale $\tau_j$ for each spin $\si _j$ and a distribution 
$W(\tau)$ of these time-scales, we obtain a theoretical prediction of the 
distribution of the number of flips per spin at time $t$:

\begin{equation}
\rho_{\rm th}(W,f,t)\simeq\int_0^\infty d\tau~W(\tau)
\lh \!\bay {c}t\\f\eay \!\rh \lh {1\ov \tau}\rh^f (1\m {1\ov \tau})^{t-f}~.
\end{equation}

\noi Minimization of $\sum_{f=0}^t\lv \rho_{\rm sim}(f,t)-\rho_{\rm
th}(W(\tau),f,t)\rv ^2$ with  respect to $W(\tau)$ then yields the most 
probable distribution of time-scales $W^*(\tau)$, see fig. 1. 

\begin{figure}[h]
\vspace*{1cm}
\setlength{\unitlength}{0.55mm}
\begin{picture}(0,90)
\put( 70,20){\epsfxsize=140\unitlength\epsfbox{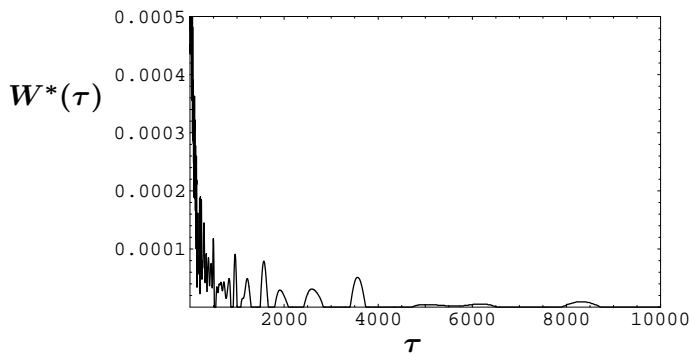}}
\put(141.3,20){\makebox(3,1){{\mbox{\small\boldmath$\tau$}}}}
\put( 55,80){\makebox(3,1){{\mbox{\small\boldmath$W^*(\tau)$}}}}
\end{picture}
\vspace*{-5mm}
\caption{The most probable distribution of time-scales for a simulation of the
SK-model with $N=6000$, after $t=5.10^5$ Monte-Carlo updates per spin, for 
$T=0.25$.}
\end{figure}

We have found that both the number of peaks (in agreement with full RSB), and 
the separation between peaks (in agreement with infinitely disparate 
time-scales) seem to grow with increasing system size and/or time. The total 
fraction of the slow spins, however, seems to remain finite, which implies that
a more precise analytical treatment for the choice of clusters or simulations
with larger system sizes and/or times are needed.

%%%%%%%%%%%%%%%%%%%%%%%%%%%%%%%%%%%%%%%%%%%%%%%%%%%%%%%%%%%%%%%%%%%%%%%%%%%%%%%%
\section{Discussion \lb {DI}}
%%%%%%%%%%%%%%%%%%%%%%%%%%%%%%%%%%%%%%%%%%%%%%%%%%%%%%%%%%%%%%%%%%%%%%%%%%%%%%%%

We have shown that the Parisi solution can be derived from simple physical
principles, and can be interpreted as describing a system with an infinite 
hierarchy of time-scales where a vanishingly small fraction of slow spins act 
as effective disorder for the faster ones. The block-sizes  $m _\ell$ at level 
$\ell$ of the Parisi matrix are found to be the ratio of the effective 
temperature $T_\ell$ of that level and the ambient temperature $T$.  
Extremization with respect to $m _\ell$ expresses the fact that for 
self-consistency the entropy of the spins slower than level $\ell$ is zero (they
are frozen on the time-scale $\ta_\ell$). It follows from physical 
considerations (i.e. the absence of heat flow in equilibrium) that
$m _\ell\leq 1~,~~\fa \ell$. The fact that the fraction of slow spins vanishes,
indicates that the cumulative entropy of the slow spins is less than extensive,
and hence that the so-called {\it complexity} is zero (at least in this full-RSB
model).

\begin{figure}[h]
\vspace*{1cm}
\setlength{\unitlength}{0.55mm}
\begin{picture}(-1,90)
%%\put(  5, 15){\makebox(3,1){\footnotesize \bf{a)}}}
\put(  5, 20){\epsfysize=80\unitlength\epsfbox{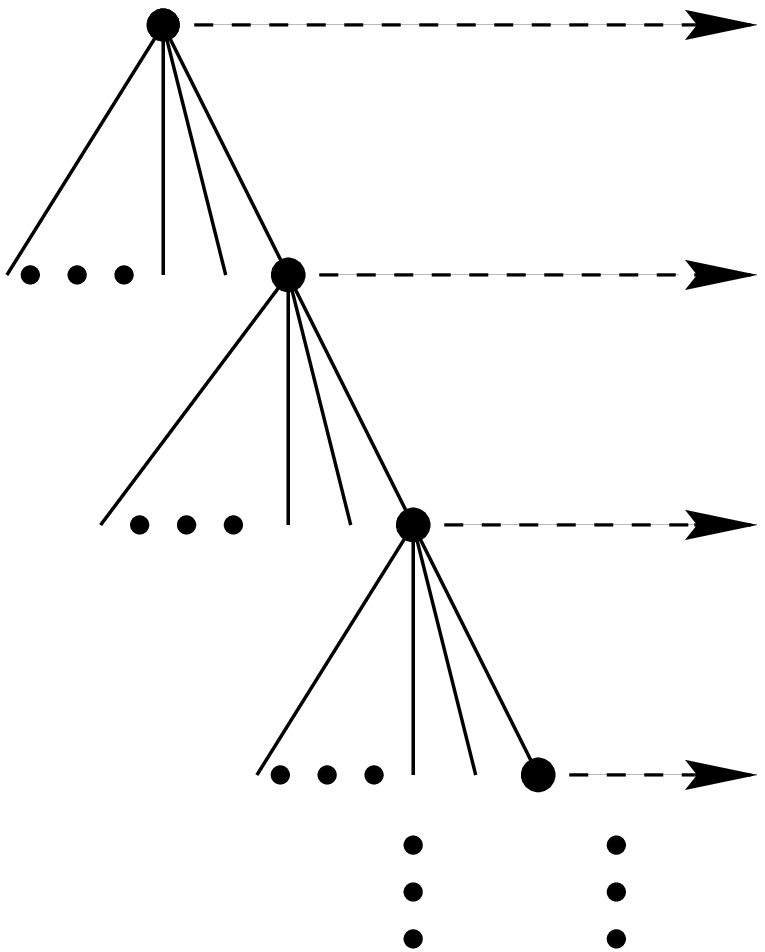}}
\put( 93,100){\makebox(3,1)
{\footnotesize same disorder                                }}
\put( 85, 79){\makebox(3,1)
{\footnotesize same \boldmath $\vec{\sigma^0}$}}
\put( 85, 58){\makebox(3,1)
{\footnotesize same \boldmath $\vec{\sigma^1}$}}
\put( 85, 37){\makebox(3,1)
{\footnotesize same \boldmath $\vec{\sigma^2}$}}
%%\put(155, 15){\makebox(3,1){\footnotesize \bf{b)}}}
\put(130, 27){\epsfxsize=140\unitlength\epsfbox{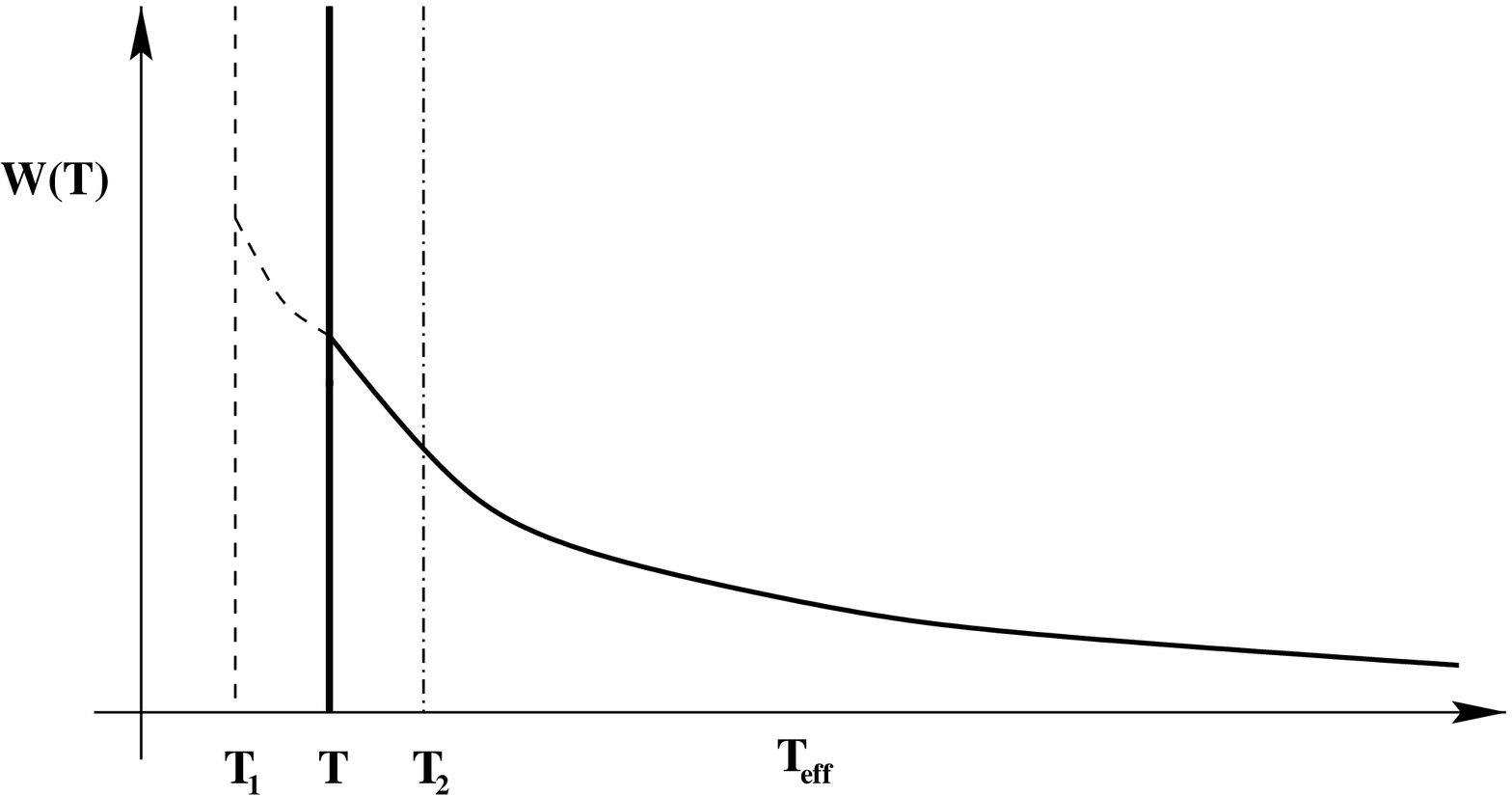}}
\end{picture}
\vspace*{-5mm}
\caption{  Left: the ultra-metric tree, which here is a direct
consequence of the hierarchy of time-scales. Right: qualitative picture of
the distribution $W(T)$ of (effective) temperatures (time-scales increase 
with $T_{\rm eff}$) at external temperature $T$ ($T_1$,$T_2$, resp.) in the 
spin glass phase.}
\end{figure}

Although a more careful treatment of the selection of clusters is obviously
required, the main consequences of our interpretation do not crucially depend on
it. Firstly, ultra-metricity (see fig. 2a) is a direct consequence of the
existence of a hierachy of time-scales. At each level $\ell$, the different
descendants of a node represent different configurations of the
$\vsi _{\ell+1}$, which, however, all share the same realisation of disorder 
and slower spins. Furthermore, a large fraction of the spins (see fig. 2b)
evolves at the fastest (microscopic) time-scale at
the ambient temperature $T$, while a small fraction of the spins evolves at
(infinitely) slower time-scales at higher effective temperatures. Therefore,
cooling to a temperature $T_1<T$ and heating back to $T$ will leave the spins
with $T_{\rm eff}>T$ unchanged, explaining memory effects. On the other
hand, after heating to $T_2>T$ and cooling back to $T$, the original
configuration of the spins with $T\leq T_{\rm eff}\leq T_2$ will be erased,
which may explain thermo-cycling experiments \cite{TC}. Note that since the 
spins are discrete variables, finding a small number of flips for a given spin 
implies long periods of stationarity (persistency) with only short periods of 
activity (avalanches).

We expect the qualitative features of our picture to survive in short range 
systems, where the time-scales need not be infinitely disparate due to 
activated processes. It is as yet unclear whether each level would represent a 
single cluster or a family of clusters. 
The origin of the slow time-scales of these clusters can only be
understood if they are coupled much stronger internally, than (effectively) to 
the rest of the system (i.e. a softened version of the completely disconnected 
clusters which give rise to so-called Griffiths singularities in diluted systems
\cite{G}). In short range systems, the clusters would have to be spatially
localised, which is in line with the {\it droplet} picture for short range spin
glasses as proposed by Fisher and Huse \cite{FH}.  The fact that the time-scale
of a clusters increases with $T_{\rm eff}-T$, explains why the effective age of
a system at a certain $T$ is found to be either older or younger when it spends
some time at a $T_1(<T)$, $T_2(>T)$ respectively.

A more careful treatment of the selection of clusters is clearly needed (and 
currently been carried out), both for full- and 1-RSB models. This may allow us 
to calculate the complexity in such systems. Furthermore, it needs to be 
investigated whether slow clusters survive above the thermodynamical (spin) 
glass temperature $T_{\rm sg}$. Our results suggest further numerical
experiments for mean field and short range models, concentrating on quantities
such as spin flip frequencies, avalanches, spatial correlations, and 
cluster persistency.

To conclude, we have shown how a scheme based on an autonomous selection of
infinitely disparate time-scales can be used succesfully to describe the 
statics of disordered systems with quenched disorder and discrete and/or 
continuous stochastic variables. It allows us to {\it derive} the Parisi scheme 
from simple physical principles, and interpret its ingredients in such a way 
that it becomes compatible with the droplet picture for short range models.
\nl 
\nl 
It is our pleasure to thank F.~Ritort and D.~Sherrington for critical comments
and stimulating discussions. 

%%%%%%%%%%%%%%%%%%%%%%%%%%%%%%%%%%%%%%%%%%%%%%%%%%%%%%%%%%%%%%%%%%%%%%%%%%%%%%%%

%%%%%%%%%%%%%%%%%%%%%%%%%%%%%%%%%%%%%%%%%%%%%%%%%%%%%%%%%%%%%%%%%%%%%%%%%%%%%%%%
%%%%%%%%%%%%%%%%%%%%%%%%%%%%%%%%%%%%%%%%%%%%%%%%%%%%%%%%%%%%%%%%%%%%%%%%%%%%%%%%
\end{document}